# SLAC UED LLRF SYSTEM UPGRADE*

L. Ma, X. Shen, K. Kim, D. Brown, M. D'Ewart, B. Hong, J. Olsen, S. Smith, D. Van Winkle, E. Williams, S. Weathersby, X. Wang, A. Young, J. Frisch
SLAC National Accelerator Laboratory, Menlo Park, CA USA

*Abstract*

The SLAC mega-electron-volt (MeV) ultrafast electron diffraction (UED) instrument is a powerful "electron camera" for the study of time-resolved, ultrafast atomic & molecular dynamics in chemical and solid-state systems. The UED laser timing synchronization is upgraded to SLAC (Advanced Telecommunications Computing Architecture) ATCA based femtosecond laser timing synchronization system. UED radio frequency (RF) gun cavity low level RF (LLRF) control is also being upgraded to SLAC ATCA based LLRF system to provide improved performance and maintainability. The new laser timing synchronization and LLRF control are implemented in one ATCA crate located in the laser room. A 2$^{nd}$ ATCA crate is set up to monitor the Klystron and sub-booster performance. With the ATCA based RF control system, the laser timing jitter has achieved 10 fs. The UED operation beam rate is being upgraded from 180 Hz to 360 Hz, which paves the way for a further upgrade of operation beam rate to 1 kHz in the future. In this paper, we present design details and characterization results of the implementation.

## INTRODUCTION

SLAC Ultrafast Electron Diffraction instrument [1-3] uses photocathode to emit electron beams which are then accelerated to Mega electron volts by a 1.6cell 2856MHz RF cavity. Highly energetic electrons take snapshots of the interior of materials when passing through them. Because electrons interact differently to materials than X ray light, UED instrument reveal different properties of materials and therefore is a complementary tool to SLAC free electron lasers.

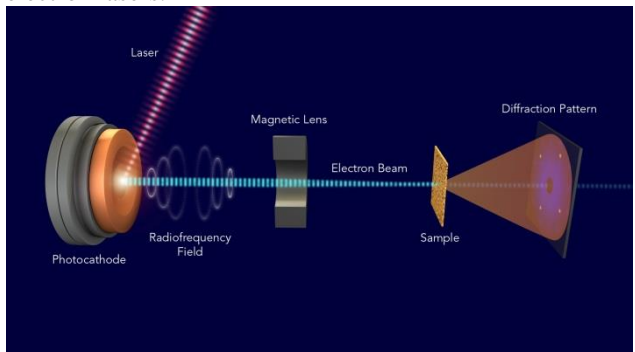

Figure 1: Schematic of a typical experimental setup of SLAC MeV UED [1].

Figure 1 shows a schematic of a typical experimental setup of SLAC MeV UED. An ultrafast (< 100 fs) ultraviolet laser triggers electron beam emission from the copper cathode in the RF gun. The laser system is seeded by a mode-locked Ti:sapphire laser oscillator. In order to get shot-by-shot synchronized and energy-stable MeV electron beam, the laser oscillator and RF system are both synchronized to a master RF reference signal. The laser oscillator outputs laser pulses at 68 MHz. The laser phase is compared with the RF reference phase to calculate a compensation signal to adjust the laser cavity length through a Piezo actuator which is mounted on a mirror in the cavity. The old laser timing synchronization system operates at 3808 MHz (56$^{th}$ harmonic of the laser pulse) rather than the 2856 MHz RF gun frequency. The old system uses separate hardware modules for RF phase detection, laser phase detection, phase comparison and an IQ phase shifter for phase set point adjustment. This system is complicated and delivers a laser timing jitter about 80 fs which cannot be further improved [2, 4]. The old UED gun cavity RF phase and amplitude control are through the PAD/PAC system [5]. Though it does provide good performance on phase/amplitude stability, the PAD/PAC system is limited at ≤ 180 Hz beam rate and cannot be upgraded to higher beam rate which is required by UED for improvement of machine performance. A new laser-and-LLRF control system is desired to overcome the above limitations. The new system also needs to fit into the existing infrastructure of the UED RF reference and timing event generation system, interlock system, laser bucket selection and coarse laser frequency tuning by motor control.

The Advanced Instrumentation for Research Division in the SLAC Technical Innovation Directorate developed a common hardware and firmware platform for beam instrumentation based on the Advanced Telecommunication Computing Architecture (ATCA). This platform incorporates fast FPGA & ADCs, multi-carrier slots, precision RF receiver, RF transmitter and backplane communication which support parallel operations of multi RF systems [6]. SLAC TID AIR RF group has developed a few beam diagnostics and control systems based on this platform [6-9]. In this paper we present the UED laser timing synchronization and gun cavity LLRF control systems upgrade based on this platform.

## DESIGN AND MEASUREMENT OF THE NEW RF CONTROL SYSTEM

The new system combines the laser timing and gun cavity control into one ATCA crate. Figure 2 shows the ATCA hardware which is installed at the UED laser room. The crate has 6 field programmable gate array (FPGA) carrier slots; each slot accommodates one RF receiver card and one RF transmitter card. Within each slot there are 10 RF down conversion channels, one RF up conversion channel and baseband channels on board. The FPGA, analog-to-digital converters (ADCs) and RF digital-to-analog converter (DAC) are clocked at 357MHz. The local oscillator (LO) frequency is 2771MHz. LO signal is provided by an on board voltage-controlled oscillator (VCO) which is locked to an external RF reference. LO locking scheme is shown in Figure. 4. The intermediate frequency (IF) is 85MHz. The ratio between IF frequency and FPGA clock is 5/21, which is chosen for best alias frequency suppression. Detailed description of the ATCA Advanced Mezzanine Card (AMC) RF cards design can be found at [7]. Figure 3 shows the schematic of the UED laser timing and the gun cavity control with the ATCA system. The two systems are implemented in separate slots so that they are independent from each other, making the maintenance for hardware/firmware/software convenient. RF reference is split into two just before the crate so that the two systems lock to the same reference.

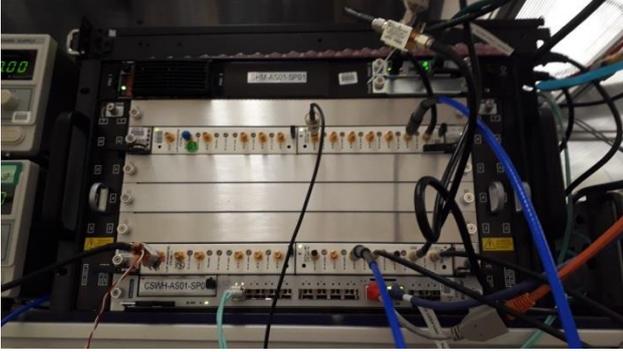

Figure 2: ATCA laser timing control and gun cavity control systems installed at SLAC UED laser room.

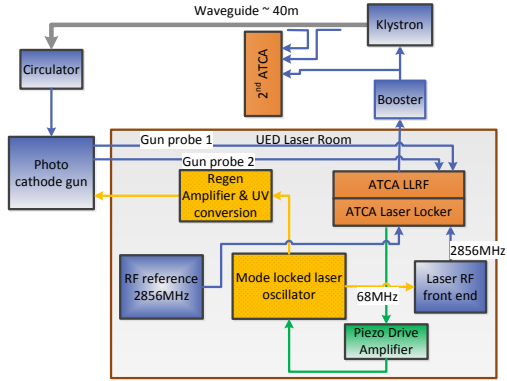

Figure 3: UED laser timing and gun cavity LLRF control system overview.

### ATCA Femtosecond Laser Timing System

A high-bandwidth (10 GHz) photodiode is used to pick up the 68 MHz train of laser pulses and convert it to electrical pulses. A low noise RF front end module designed in house is used to pick up the 2856 MHz signal which is the 42th harmonic of the laser pulse. The pickup signal which contains the laser phase information is sent to ATCA for processing. Figure 4 shows the algorithm of the laser timing control. The RF reference and laser pick up are read in by ADC and processed in parallel to get their phases. The phase error between laser and RF reference goes through a proportional-integral (PI) controller and low pass filter to get the phase compensation signal which is used to adjust the laser cavity length through a Piezo actuator. The laser Piezo tuning range is about 15 kHz with a tuning gain of 1.5 kHz/volt with respect to the ATCA baseband DAC which has a ±5 Volts output range. A low noise amplifier is designed in house to amplify the DAC output to 100 Volts to drive the Piezo actuator. The DAC can output as fast as several MHz frequency signal. The laser Piezo actuator has a charge capacitance around 1 μs. The Piezo amplifier has a 5 Ω output resistance. The bandwidth of the Piezo actuator is calculated below.

$$f_c = \frac{1}{2\pi RC} = 32 \; kHz. \quad (1)$$

It turns out that the Piezo actuator is the limitation of the system response speed. The open loop transfer function of the system is given below.

$$H = K_p(1+\frac{K_i}{s})^2 \frac{\omega_{lp}}{s+\omega_{lp}} \frac{K_{Piezo}}{s}, \quad (2)$$

where $K_{Piezo}$ is the Piezo tuning gain, $K_p$, $K_i$ and $\omega_{lp}$ are the loop proportional gain, integrational gain and bandwidth of the low pass filter respectively. Second order integration is implemented to suppress low frequency phase noise. By carefully tuning the photo diode pickup position and power coupling level to make sure the photo diode is in its linear response region and by optimizing the loop control parameters, we have achieved 20 kHz loop bandwidth and 10 femtosecond laser timing jitter. Figure 5 shows the laser phase noise measurement and timing jitter calculation.

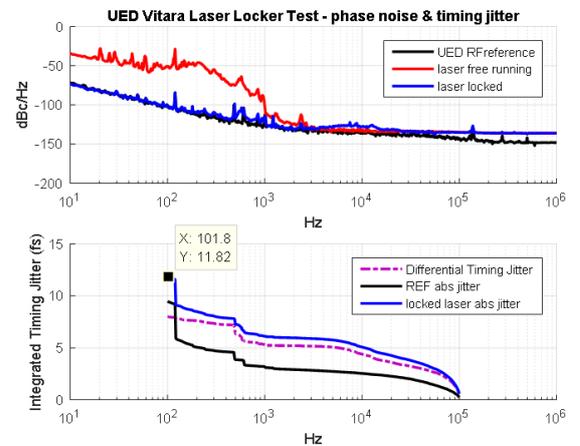

Figure 5: UED laser phase noise measurement and timing jitter calculation.

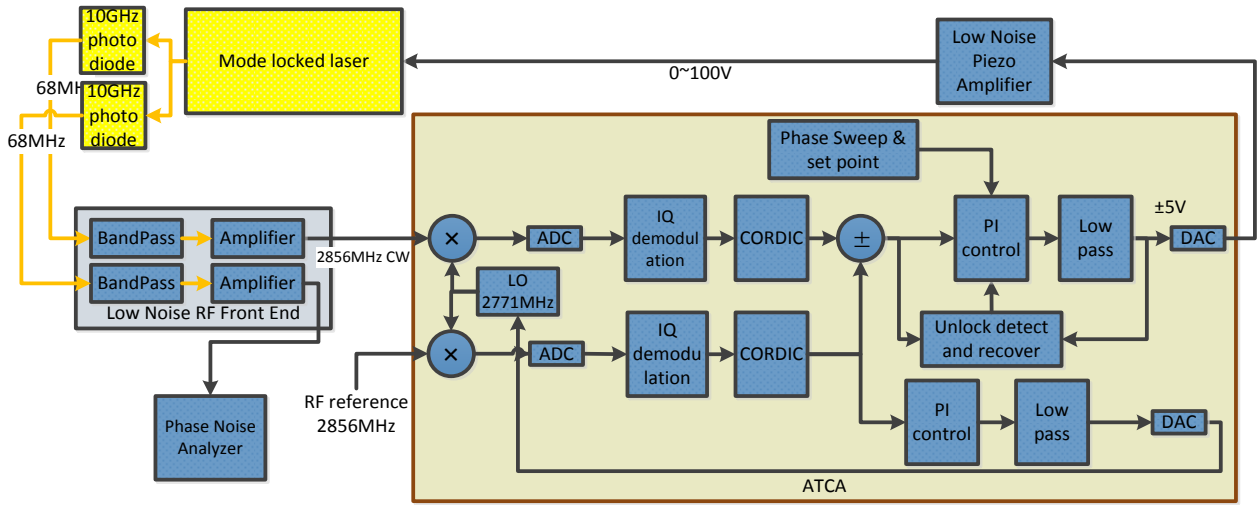

Figure 4: ATCA based Femtosecond laser timing synchronization algorithm

To measure the laser phase noise, a 2nd laser diode is used to couple the laser pulse out and the signal is sent through the 2nd channel of the RF front end module and converted to CW 2856 MHz signal. The RF reference phase noise (black curve) is also measured for comparison. The laser exhibits large phase noise (red curve) in low frequency (f<1 kHz) region when running free. When the feedback loop is closed, the laser phase noise (blue curve) drops down to be close to the RF reference phase noise level. In low frequency region, the laser noise level follows the RF reference phase noise level closely. Laser phase noise is much lower than RF reference phase noise at high frequency (f>100 kHz). Therefore in the timing jitter calculation, we have integrated phase noise from 100 Hz to 100 kHz. In order to see the high frequency contribution to phase noise clearly, we have carried out timing jitter integration from high frequency to low frequency. Figure 5 shows that the RF reference integrated timing jitter is 10 fs, the laser integrated timing jitter is 11 fs, the differential timing jitter between laser and RF reference is 8 fs. Compared to the old laser timing system, new system significantly improves the laser timing synchronization.

## ATCA LLRF System

The ATCA based LLRF system was first developed for LCLS Mission Readiness project. Detailed description of the system can be found in [7-8]. UED gun cavity LLRF control system upgrade is mostly similar to the Mission Readiness project except that the beam rate is increased from 180 Hz to 360 Hz and UED is planning to further increase beam rate for kHz operation in the future. UED RF pulse length is 2μs or less, which is comparable to the RF system loop delay. Because of the short RF pulse length and low beam rate, the amplitude and phase control is designed to be pulse to pulse feedback with a very low integration gain. Two gun probe pickups are used in the feedback control with weighing factor 50% each. Because of the low beam rate, the PI controller is implemented in software. Figure 6 shows the gun cavity LLRF control algorithm.

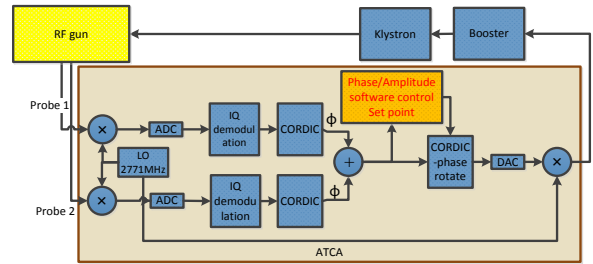

Figure 6: Gun cavity control algorithm

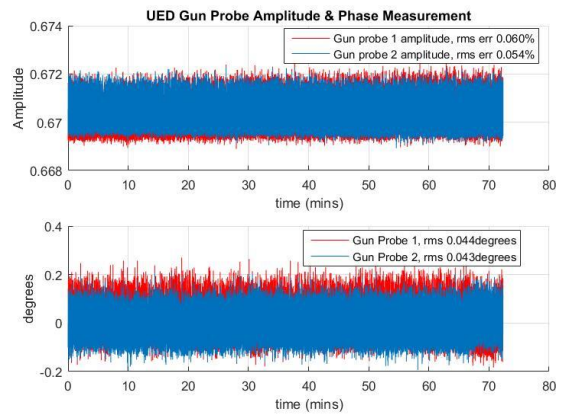

Figure 7: Gun probe amplitude & phase measurement

In order to cancel out the 60 Hz resonance on cavity phase noise spectrum, the amplitude and phase feedback are implemented in 6 individual loops at 60 Hz operation rate. Figure 7 shows the gun probe amplitudes & phases measured in 70 minutes. Data collection sample rate is 60Hz. Rms phase error of gun probes are about 0.04 de-grees. Amplitude rms errors are 0.06% and 0.054%. Figure 8 shows the differential phase error between the two gun probes and the differential phase jitter integrated from 30 Hz to 0.2 mHz. Two 5m Helix cables are used to connect gun probes to ATCA inputs. Subtracting the phase noise of the two probes cancels out the common mode noise of the two channels. The differential phase noise between the two channels is 0.02 degrees.

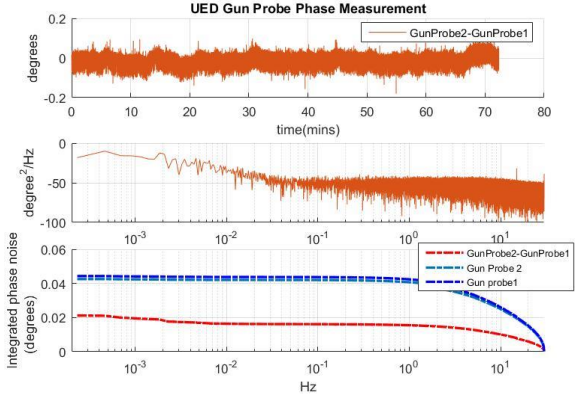

Figure 8: Differential phase error between two gun probes

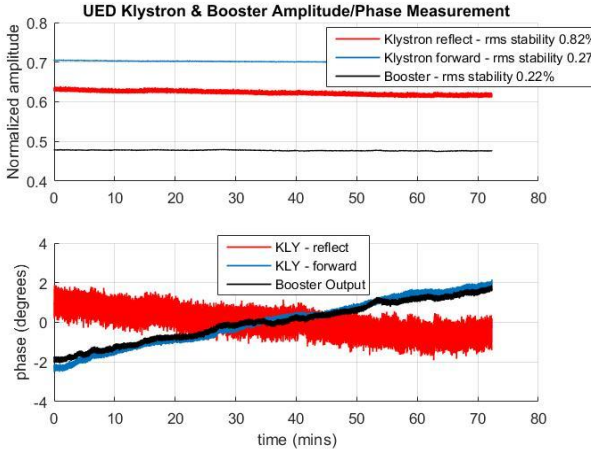

Figure 9: Klystron and Booster Amplitude/phase measurement

## 2$^{nd}$ ATCA System at Klystron Area

UED Klystron and Booster are located 40 meters away from the gun cavity. Monitoring the klystron and booster power level, amplitude/phase drifting at beam operation rate is desirable. Therefore a 2$^{nd}$ ATCA system is installed into the booster rack so that booster output, klystron forward and reflected signals can be easily measured. Figure 9 shows measurements taken at 60 Hz sample rate. The design of the LLRF system has enabled the measurement to be taken at specific time within the short RF pulse. Figure 9 shows that klystron forward amplitude rms noise is 0.27% and booster amplitude rms noise is 0.22%. The klystron reflected amplitude has a much bigger amplitude noise 0.82%. Both klystron and booster phases have large drifting range ±2degrees within 70 minutes The much-less-stable phase and amplitude performance at the booster and the klystron is due to the response of the LLRF control system to the environmental change in order to maintain the stable RF in the gun cavity.

## Electron Beam Energy Measurement

In a UED experiment, a change in the electron beam energy results in a change in the electron beam arrival time to the sample, which can smear out the ultrafast transient structural information during the data accumulation process. Therefore, maintaining stable electron beam energy is substantial to achieve a high time resolution. We have carried out direct single-shot electron beam energy measurement to characterize the electron beam energy stability controlled by the ATCA system. In the measurement, electron beam passed through a spectrometer and was bent onto a detector. Panel (a) in Fig. 10 shows an example of the two-dimensional (2d) profile of the bent electron beam on the detector. Since the bending occurs in the horizontal direction, the horizontal centroid of the beam profile is linearly correlated to the electron energy. We obtained a time trace of electron beam energy over 70 minutes, as shown in panel (c) of Fig. 10. Panel (d) in Fig. 10 shows a histogram of the electron beam energy distribution. The rms energy spread was estimated from a Gaussian fit to be 0.023%, which is estimated to cause an rms electron beam arrival time jitter as low as 12 fs.

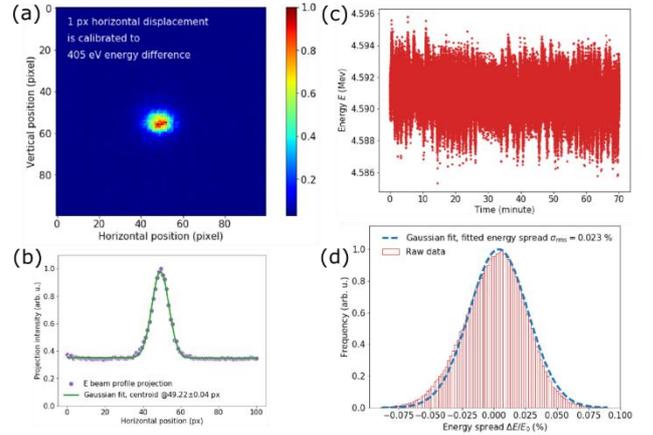

Figure 10: Electron beam energy stability characterization results. (a) A typical single-shot two-dimensional electron beam profile after the electron beam passes through a spectrometer. (b) Electron beam profile projected onto the horizontal axis corresponding to the case in panel (a). The projection was fitted to a Gaussian profile to determine its centroid position from which the electron beam energy was deduced. (c) Measured single-shot electron beam energies over 70 minutes. (d) Histogram of the electron beam energy distribution corresponding to the case in panel (c). The rms energy spread was estimated from a Gaussian fit to be 0.023%.

## CONCLUSION

We have designed and commissioned an ATCA based femtosecond laser timing control and gun cavity control system for SLAC MeV UED instrument. The UED rms laser timing jitter is reduced to 10 fs. UED beam operation rate is increased to 360 Hz. The new system has the potential to support kHz beam operation rate. Gun cavity rms amplitude jitter is controlled to about 0.05%. Gun cavity rms phase jitter is controlled to 0.04 degrees. Direct measurement of electron beam energy regulated by the new ATCA system shows a 0.023% energy stability, which is estimated to cause only 12 fs electron beam arrival time jitter. The ATCA hardware itself has an amplitude noise 0.01% and phase noise 0.01degrees at 3GHz [7]. Currently the ATCA crate is 5 meters away from the gun probe pick up point. By reducing this distance, we expect better amplitude and phase noise control.